\documentclass[conference]{IEEEtran}
\usepackage{listings}
\usepackage{xcolor}

\definecolor{keywords}{RGB}{255,0,90}
\definecolor{comments}{RGB}{0,0,113}
\definecolor{red}{RGB}{160,0,0}
\definecolor{green}{RGB}{0,150,0}
\title{FSK Demodulation and Bit String Extraction: A Python-Centric Approach in SDR Systems}

\author{\IEEEauthorblockN{Jimi Sanchez}
\IEEEauthorblockA{
jimi.c.sanchez@nasa.gov\\
jimi.linuxguy@gmail.com\\
https://jimisanchez.com
}
}
\date{February 2024}
\begin{document}
\maketitle

\begin{abstract}
    Frequency Shift Keying (FSK) modulation is widely utilized in various communication systems for data transmission due to its simplicity and robustness. In this paper, we present a Python-centric approach for demodulating FSK signals and extracting bit strings in Software Defined Radio (SDR) systems. Leveraging the flexibility and power of Python programming language along with SDR platforms, we explore the intricacies of FSK demodulation techniques and efficient bit string extraction methods. Our approach focuses on real-time processing capabilities, enabling rapid decoding of FSK signals with minimal latency. We discuss the implementation details, performance considerations, and optimization strategies, highlighting the advantages and challenges of utilizing Python in SDR applications. Furthermore, we demonstrate the effectiveness of our approach through experimental results and comparisons with existing methods. This paper serves as a comprehensive guide for researchers and practitioners interested in implementing FSK demodulation and bit string extraction algorithms using Python within the context of SDR systems.
\end{abstract}

\section{Introduction}
Software Defined Radio (SDR) has revolutionized the field of wireless communication by offering unparalleled flexibility, enabling the implementation of diverse modulation and demodulation techniques with software-based algorithms. Within this realm, Frequency Shift Keying (FSK) modulation stands out as a popular method for data transmission due to its simplicity, efficiency, and resistance to noise \cite{knoedler2021}.

This paper focuses on the decoding of FSK signals into bit strings using Python within the context of SDR systems. While FSK modulation is relatively straightforward, demodulating these signals and accurately extracting the transmitted information pose challenges, particularly in real-time scenarios. Leveraging the power of Python, a versatile and widely-used programming language, in conjunction with SDR platforms, offers a compelling solution to address these challenges.

In this paper, we delve into the intricacies of FSK demodulation and bit string extraction, presenting a Python-centric approach that combines signal processing techniques with SDR capabilities. We explore various demodulation algorithms and bit string extraction methods, discussing their implementation details and performance considerations. Furthermore, we investigate optimization strategies to enhance the efficiency and reliability of FSK signal decoding in Python-based SDR systems.

Through experimental validation and comparative analysis, we demonstrate the effectiveness of our approach and highlight its practical implications for wireless communication applications. By providing insights into FSK demodulation using Python in SDR environments, this paper aims to facilitate further research and development in the field, fostering innovation and advancement in wireless communication technologies.
\section{Background: FSK Modulation and Demodulation}
Frequency Shift Keying (FSK) modulation is a digital modulation technique widely employed in telecommunications for transmitting digital data over radio frequency (RF) channels. In FSK modulation, digital data is represented by alternating the frequency of the carrier signal between two or more predetermined frequencies, known as "mark" and "space" frequencies, corresponding to binary states.

The simplicity and robustness of FSK modulation make it a popular choice for various applications, including wireless communication systems, RFID (Radio Frequency Identification) systems, and telemetry. FSK modulation offers advantages such as resistance to noise, efficient use of bandwidth, and straightforward implementation.

In an FSK modulation scheme, the modulated signal's frequency shifts between discrete frequency values, typically referred to as the "mark" and "space" frequencies. The frequency deviation between these two frequencies determines the modulation index and influences the modulation's bandwidth efficiency and spectral characteristics.

Demodulating FSK signals involves recovering the original digital data from the modulated carrier signal. Various demodulation techniques exist for FSK signals, including coherent and non-coherent demodulation methods. Coherent demodulation relies on knowledge of the carrier signal's phase and frequency, whereas non-coherent demodulation does not require phase synchronization.

Once the FSK signal is demodulated, the next step is to extract the transmitted bit string accurately. This process involves thresholding the demodulated signal to discriminate between the "mark" and "space" frequencies and recovering the digital data symbols.

In recent years, the advent of Software Defined Radio (SDR) technology has transformed the landscape of wireless communication. SDR systems offer the flexibility to implement modulation and demodulation techniques in software, enabling rapid prototyping, experimentation, and deployment of new communication protocols and algorithms.

Python has emerged as a prominent programming language for signal processing and SDR applications due to its ease of use, extensive libraries (such as NumPy and SciPy), and vibrant community support. By leveraging Python in conjunction with SDR platforms, researchers and practitioners can develop efficient and scalable solutions for demodulating FSK signals and extracting bit strings in real-world scenarios.

In this paper, we explore the challenges and opportunities in decoding FSK signals into bit strings using Python within the context of SDR systems. We investigate various demodulation techniques, bit string extraction methods, implementation details, performance considerations, and optimization strategies, aiming to provide insights and guidance for researchers and practitioners working in the field of wireless communication and software-defined radio.
\section{Overview of Software Defined Radio (SDR) Systems}
Software Defined Radio (SDR) systems have revolutionized the field of wireless communication by shifting the paradigm from hardware-centric to software-centric radio architectures. Unlike traditional radio systems, where the functionality is largely determined by dedicated hardware components, SDR systems leverage software to perform many of the radio signal processing tasks, offering unprecedented flexibility, reconfigurability, and scalability.

At the core of an SDR system is a general-purpose processor or a specialized digital signal processing (DSP) unit, which executes signal processing algorithms to perform modulation, demodulation, filtering, encoding, decoding, and other radio-related functions. The hardware platform typically consists of an analog-to-digital converter (ADC) for digitizing the incoming RF signal and a digital-to-analog converter (DAC) for generating the outgoing RF signal.

One of the key advantages of SDR systems is their ability to support multiple communication standards and protocols through software reconfiguration, eliminating the need for dedicated hardware for each standard. This enables SDR platforms to adapt dynamically to changing requirements, upgrade to new standards, and accommodate emerging technologies with minimal hardware modifications.

SDR systems are widely used in various applications, including military communications, public safety networks, amateur radio, cognitive radio, spectrum monitoring, and research and development. They offer flexibility in frequency band selection, modulation schemes, channel bandwidths, and signal processing algorithms, making them ideal for experimentation, prototyping, and rapid deployment of new radio systems.

The development of SDR software is facilitated by a rich ecosystem of programming languages, libraries, and frameworks. Python has emerged as a popular choice for SDR development due to its simplicity, readability, extensive libraries (such as GNU Radio, SoapySDR, pyrtlsdr), and community support. Python enables rapid prototyping and implementation of complex signal processing algorithms, making it well-suited for SDR applications.

In this paper, we leverage the capabilities of SDR systems and Python programming to decode Frequency Shift Keying (FSK) signals into bit strings. By combining the flexibility of SDR platforms with the versatility of Python, we aim to provide insights and methodologies for efficiently implementing FSK demodulation and bit string extraction in software-defined radio environments.
\section{Python in SDR: A Powerful Combination}
\section{FSK Demodulation Techniques}
Frequency Shift Keying (FSK) demodulation is the process of recovering the original digital data from the modulated FSK signal. Several demodulation techniques exist, each with its advantages and considerations. In this section, we discuss three primary FSK demodulation techniques: Coherent Demodulation, Non-Coherent Demodulation, and Differential Demodulation.

Coherent Demodulation:
Coherent demodulation is a technique that requires knowledge of the carrier signal's phase and frequency. It involves generating a local oscillator at the same frequency as the carrier signal and maintaining phase synchronization between the received signal and the local oscillator. The received signal is mixed with the local oscillator to downconvert it to baseband, followed by low-pass filtering to remove high-frequency components. Finally, the baseband signal is sampled and processed to extract the digital data.
Advantages:

Provides optimal demodulation performance in ideal conditions.
Suitable for situations where carrier phase and frequency information are available.
Considerations:

Requires accurate carrier phase and frequency estimation.
Susceptible to phase noise and frequency offset errors, especially in practical scenarios.
Non-Coherent Demodulation:
Non-coherent demodulation is a technique that does not require phase synchronization between the received signal and the demodulator. Instead, it relies on envelope detection or magnitude estimation to recover the digital data. In FSK modulation, non-coherent demodulation involves squaring the received signal to obtain its instantaneous power, followed by low-pass filtering to extract the envelope. The envelope is then sampled and processed to determine the transmitted symbols.
Advantages:

Robust to phase noise and frequency offset errors.
Suitable for applications where carrier phase synchronization is challenging.
Considerations:

May suffer from degraded performance compared to coherent demodulation, especially in low signal-to-noise ratio (SNR) conditions.
Susceptible to amplitude variations and fading effects.
Differential Demodulation:
Differential demodulation is a technique that exploits the phase difference between consecutive symbols rather than absolute phase information. In FSK modulation, differential demodulation involves comparing the phase or frequency of successive symbols to determine the transmitted bit. This technique eliminates the need for accurate carrier phase estimation and synchronization.
Advantages:

Robust to phase and frequency offset errors.
Suitable for low-complexity demodulation implementations.
Considerations:

May require additional error correction mechanisms to mitigate error propagation.
Performance depends on the choice of differential encoding scheme and symbol timing synchronization.
In practice, the choice of FSK demodulation technique depends on various factors such as the signal-to-noise ratio (SNR), carrier frequency stability, phase noise characteristics, hardware complexity, and computational resources. By understanding the principles and characteristics of different demodulation techniques, researchers and practitioners can select the most appropriate approach for their specific application requirements.
\section{Bit String Extraction Methods}
Once the FSK signal has been demodulated, the next crucial step is to extract the transmitted bit string accurately. This process involves converting the demodulated signal into a sequence of binary symbols representing the original digital data. Several methods can be employed for bit string extraction, each with its advantages and considerations. In this section, we discuss three primary bit string extraction methods: Thresholding, Timing Recovery, and Error Correction Techniques.

Thresholding:
Thresholding is a straightforward method for extracting the bit string from the demodulated FSK signal. It involves setting a threshold value and comparing the demodulated signal's amplitude or power against this threshold. Depending on whether the signal amplitude is above or below the threshold, the corresponding binary symbol (0 or 1) is determined \cite{smith2020}.
Advantages:

Simple and computationally efficient.
Suitable for scenarios with relatively clean and noise-free demodulated signals.
Considerations:

Susceptible to noise and amplitude variations, which can degrade performance, especially in low signal-to-noise ratio (SNR) conditions.
May require careful selection of the threshold value to balance between false positive and false negative detections.
Timing Recovery:
Timing recovery is a method for synchronizing the sampling instants with the symbol boundaries in the demodulated signal. In FSK modulation, accurate symbol timing synchronization is essential for correct bit string extraction. Timing recovery algorithms estimate the symbol timing offset and adjust the sampling instants accordingly to maximize the signal-to-noise ratio (SNR) and minimize inter-symbol interference (ISI).
Advantages:

Improves the accuracy and reliability of bit string extraction, especially in the presence of timing jitter and symbol timing offset.
Mitigates ISI effects and improves detection performance.
Considerations:

Requires knowledge of the symbol timing characteristics and synchronization algorithms.
Computational complexity increases with the symbol rate and timing recovery accuracy requirements.
Error Correction Techniques:
Error correction techniques aim to improve the robustness of bit string extraction by detecting and correcting errors introduced during transmission. Forward Error Correction (FEC) codes, such as convolutional codes and Reed-Solomon codes, are commonly used for error detection and correction in digital communication systems. These codes add redundancy to the transmitted data, allowing the receiver to detect and correct a certain number of errors.
Advantages:

Enhances the reliability of bit string extraction, especially in noisy or error-prone communication channels.
Provides a means to achieve a desired level of error performance and trade-off between data rate and error correction capability.
Considerations:

Increases the overhead (redundancy) of the transmitted data, reducing the effective data rate.
Requires additional computational resources for encoding and decoding FEC codes.
In practice, a combination of these bit string extraction methods may be employed to achieve the desired level of performance and reliability. By understanding the characteristics and trade-offs of different extraction methods, researchers and practitioners can design robust and efficient FSK demodulation systems for various communication applications.
\section{Implementation Details}
For our experiment, we implemented the FSK demodulation and bit string extraction algorithms using Python and relevant signal processing libraries. The implementation was carried out on a standard desktop computer equipped with a multi-core CPU and sufficient memory.

First, we utilized the NumPy library to generate a sample FSK signal for testing purposes. The signal was generated by modulating a sequence of digital data symbols onto a carrier frequency using the specified modulation index and symbol duration.

Next, we employed the SciPy library to perform FSK demodulation on the generated signal. Depending on the demodulation technique under investigation (coherent, non-coherent, or differential), we implemented the corresponding demodulation algorithm. For coherent demodulation, we estimated the carrier frequency and phase using a phase-locked loop (PLL) and applied a mixer to downconvert the signal to baseband. Non-coherent demodulation involved envelope detection followed by low-pass filtering to extract the envelope. Differential demodulation compared the phase or frequency of successive symbols to determine the transmitted bit.

Following demodulation, we applied thresholding to the demodulated signal to extract the bit string. A threshold value was determined based on the signal's amplitude characteristics, and each sample in the demodulated signal was compared against this threshold to determine the corresponding bit (0 or 1).

Additionally, we implemented timing recovery algorithms to synchronize the sampling instants with the symbol boundaries in the demodulated signal. Timing recovery involved estimating the symbol timing offset and adjusting the sampling instants accordingly to maximize the signal-to-noise ratio (SNR) and minimize inter-symbol interference (ISI).

To evaluate the performance of our implementation, we conducted extensive experiments using simulated and real-world FSK signals. We measured key performance metrics such as bit error rate (BER), signal-to-noise ratio (SNR), and throughput. The experiments were conducted under various conditions to assess the robustness and efficiency of our FSK demodulation and bit string extraction algorithms.

Overall, our implementation provided a comprehensive framework for decoding FSK signals into bit strings using Python. By leveraging the capabilities of Python libraries and signal processing techniques, we demonstrated the feasibility and effectiveness of our approach for FSK demodulation in software-defined radio systems.

\section{Experimental Results}
Our experiments yielded insightful results regarding the performance and effectiveness of the implemented FSK demodulation and bit string extraction algorithms. We evaluated the performance of our approach under various conditions, including different signal-to-noise ratios (SNRs), modulation indices, and symbol rates.

First, we assessed the accuracy of bit string extraction under varying SNR levels. We observed that the bit error rate (BER) increased as the SNR decreased, indicating the impact of noise on the demodulated signal's integrity. However, our implementation demonstrated robust performance even at low SNR levels, thanks to the utilization of thresholding and error correction techniques.

Next, we evaluated the efficiency of timing recovery algorithms in synchronizing the sampling instants with the symbol boundaries. We observed that accurate symbol timing synchronization significantly improved the BER and signal recovery performance, particularly in the presence of timing jitter and symbol timing offset.

Furthermore, we compared the performance of different demodulation techniques (coherent, non-coherent, and differential) in terms of BER, computational complexity, and robustness to noise. Our results indicated that coherent demodulation achieved the lowest BER under ideal conditions but was susceptible to phase noise and frequency offset errors. Non-coherent demodulation exhibited robust performance in noisy environments but at the cost of increased BER. Differential demodulation provided a balance between performance and complexity, offering robustness to phase and frequency offset errors with moderate computational overhead.

Additionally, we assessed the throughput of our implementation in processing FSK signals in real-time. We measured the processing time required for demodulation and bit string extraction and compared it against the symbol rate to determine the system's efficiency and scalability.

Overall, our results demonstrate the effectiveness of our Python-based approach for decoding FSK signals into bit strings in software-defined radio systems. By leveraging signal processing techniques and Python libraries, we achieved robust and efficient FSK demodulation, paving the way for enhanced communication systems and applications.
\section{Practical Considerations and Challenges}
While implementing FSK demodulation and bit string extraction using Python in SDR systems offers numerous advantages, several practical considerations and challenges must be addressed to ensure successful deployment and operation. In this section, we discuss some of the key practical considerations and challenges:

Computational Resources: Python-based signal processing algorithms may impose significant computational demands, especially when processing high-speed or high-volume data streams. Adequate computational resources, such as CPU cores and memory, are essential to ensure real-time performance and responsiveness.

Real-Time Processing: Achieving real-time processing capabilities in Python can be challenging due to the language's inherent interpreted nature and potential overheads associated with dynamic memory management and garbage collection. Careful optimization and profiling are required to meet stringent latency requirements.

Hardware Compatibility: Python-based SDR applications must interface seamlessly with hardware components, such as radio transceivers and analog-to-digital converters (ADCs). Ensuring compatibility and driver support for a wide range of hardware platforms and peripherals is crucial for interoperability and versatility.

Signal Integrity and Quality: Maintaining signal integrity and quality throughout the processing chain is paramount for accurate demodulation and reliable bit string extraction. Signal conditioning, filtering, and noise reduction techniques may be necessary to enhance signal-to-noise ratio (SNR) and mitigate interference effects.

Synchronization and Timing: Synchronization and timing issues can significantly impact the performance of FSK demodulation systems, particularly in multi-channel or multi-user environments. Ensuring accurate symbol timing synchronization and carrier frequency alignment is essential to avoid symbol timing offset and frequency drift.

Robustness to Channel Variability: FSK demodulation algorithms must be robust to channel impairments and variations, such as multipath fading, Doppler effects, and frequency selective fading. Adaptive algorithms and error correction techniques may be employed to mitigate the impact of channel variability and enhance system robustness.

Error Handling and Recovery: Effective error handling and recovery mechanisms are essential to ensure reliable operation in the presence of transmission errors or packet loss. Forward error correction (FEC) codes, retransmission protocols, and error detection algorithms can help detect and correct errors in the received bit stream.

Security Considerations: Security considerations, such as encryption, authentication, and data integrity, are critical for protecting sensitive information transmitted over the air. Implementing robust security mechanisms and encryption protocols is essential to safeguard against unauthorized access and malicious attacks.

\section{Conclusion and Future Directions}
In conclusion, our study has demonstrated the feasibility and effectiveness of using Python for decoding Frequency Shift Keying (FSK) signals into bit strings within the context of Software Defined Radio (SDR) systems. Through the implementation and evaluation of FSK demodulation and bit string extraction algorithms, we have shown that Python, coupled with signal processing libraries, offers a powerful platform for developing robust and efficient communication systems.

We have investigated various FSK demodulation techniques, including coherent, non-coherent, and differential demodulation, and evaluated their performance in terms of bit error rate (BER), computational complexity, and robustness to noise. Additionally, we have explored different methods for bit string extraction, such as thresholding, timing recovery, and error correction techniques, and assessed their impact on the accuracy and reliability of the decoded bit stream.

Our experiments have provided valuable insights into the strengths and limitations of each demodulation technique and bit string extraction method, facilitating informed decision-making in the design and implementation of FSK demodulation systems. Furthermore, we have demonstrated the scalability and flexibility of our Python-based approach, enabling rapid prototyping and experimentation in diverse communication scenarios.

Future Work:

While our study has provided a solid foundation for FSK demodulation using Python in SDR systems, several avenues for future research and development remain open:

Optimization and Performance Enhancement: Investigate techniques for optimizing the computational performance and throughput of FSK demodulation algorithms in Python, particularly for real-time processing applications.

Adaptive Demodulation: Explore adaptive demodulation algorithms that can dynamically adjust parameters based on the signal characteristics and environmental conditions to improve robustness and efficiency.

Integration with Hardware: Investigate integration with hardware-accelerated platforms, such as field-programmable gate arrays (FPGAs) or graphics processing units (GPUs), to leverage parallel processing and hardware resources for enhanced performance.

Advanced Error Correction: Explore advanced error correction techniques, such as turbo codes or low-density parity-check (LDPC) codes, to improve the resilience of FSK demodulation systems to channel impairments and noise.

Compatibility with Other Modulation Schemes: Extend the implementation to support other modulation schemes, such as phase shift keying (PSK) or quadrature amplitude modulation (QAM), for comprehensive SDR applications.

By addressing these areas of future work, we can further enhance the capabilities and applicability of Python-based FSK demodulation systems in SDR environments, contributing to the advancement of wireless communication technologies.

\appendix
This code snippet generates a sample FSK signal with two frequencies (frequency\_1 and frequency\_2) and then demodulates it using Fourier transform. The resulting demodulated signal is then thresholded to extract the bit string.

\begin{lstlisting}[
    linewidth=\columnwidth,
language=Python,
basicstyle=\ttfamily\tiny, 
keywordstyle=\color{keywords},
commentstyle=\color{comments},
stringstyle=\color{red},
showstringspaces=false,
identifierstyle=\color{green},
breaklines=true
]
import numpy as np

# Define FSK demodulation parameters
sampling_rate = 1000  # Sampling rate in Hz
frequency_1 = 10  # Frequency of the first FSK symbol in Hz
frequency_2 = 20  # Frequency of the second FSK symbol in Hz
symbol_duration = 1  # Duration of each FSK symbol in seconds
samples_per_symbol = symbol_duration * sampling_rate

# Generate a sample FSK signal (you would replace this with your actual input signal)
t = np.linspace(0, symbol_duration, samples_per_symbol, endpoint=False)
fsk_signal = np.concatenate((np.sin(2 * np.pi * frequency_1 * t), np.sin(2 * np.pi * frequency_2 * t)))

# Perform FSK demodulation
demodulated_signal = np.abs(np.fft.fft(fsk_signal))

# Thresholding to extract bit string
threshold = np.mean(demodulated_signal)
bit_string = "".join("1" if x > threshold else "0" for x in demodulated_signal)

# Print the extracted bit string
print("Extracted bit string:", bit_string)

\end{lstlisting}

\end{document}